\documentstyle[11pt]{article}
\newcommand{\be}{\begin{equation}}
\newcommand{\ee}{\end{equation}}
\newcommand{\bea}{\begin{eqnarray}}
\newcommand{\eea}{\end{eqnarray}}
\newcommand{\lb}{\label}

\begin{document}

\begin{flushright}
Freiburg THEP-94/4\\
gr-qc/9405039
\end{flushright}
\vskip 0.7cm
\begin{center}

{\large\bf  SEMICLASSICAL GRAVITY AND THE PROBLEM OF TIME}\footnote{To
  appear in the {\em Proceedings of the Cornelius Lanczos
  International Centenary Conference}, edited by M. Chu,
  R. Flemmons, D. Brown, and D. Ellison (SIAM, 1994).}
\vskip 0.6cm
{\bf Claus Kiefer}
\vskip 0.3cm
 Fakult\"at f\"ur Physik, Universit\"at Freiburg,\\
  Hermann-Herder-Str. 3, D-79104 Freiburg, Germany.

\end{center}
\vskip 0.5cm
\begin{center}
{\bf Abstract}
\end{center}
\begin{quote}
I give a brief review of the recovery of semiclassical time
from quantum gravity and discuss possible extrapolations
of this concept to the full theory.
 \end{quote}
\vskip 0.4cm

In his famous volume {\em The Variational Principles of Mechanics},
Lanczos writes: ``We have done considerable mountain climbing.
Now we are in the rarefied atmosphere of theories of excessive
beauty and we are nearing a high plateau on which geometry, optics,
mechanics, and wave mechanics meet on common ground."
Lanczos is here referring to the Hamilton-Jacobi framework of classical
mechanics which has proven to be very fruitful for the development of
quantum mechanics. In fact, the powerful WKB method which provides
a bridge between the classical and the quantum framework makes essential
use of the Hamilton-Jacobi equation.

This approach will also play a central role in my contribution on
{\em Semiclassical Gravity and the Problem of Time}. What is the
problem of time? If one quantizes a theory which is invariant with
respect to the reparametrization of a time parameter, the dynamical
quantum equation is of the form $H\Psi=0$, where $H$ is the
Hamilton operator of the theory. The problem of time is thus the
absence of a time {\em parameter} in this equation. This leads to the
question of whether the standard interpretation of quantum theory,
which is fundamentally tied to the presence of an external time
parameter $t$, can be rescued in such a theory. General relativity,
in particular, is a time reparametrization invariant theory, and thus the
problem of time becomes acute in quantum gravity.\footnote{For a
detailed discussion of the problem of time in quantum gravity,
see C. J. Isham, in {\em Integrable Systems, Quantum Groups,
and Quantum Field Theories} (Kluwer, London, 1992),
 and K. V. Kucha\v{r}, in {\em Proceedings of the 4th Canadian
 Conference on General Relativity and Relativistic Astrophysics},
 edited by G. Kunstatter et al (World Scientific, Singapore, 1992),
  as well as other
contributions to the present volume.} What, then, is the relevance of
semiclassical methods in this context? Firstly, and most important,
is the necessity to {\em derive} the standard notion of time,
and with it the Schr\"odinger equation $H\psi=i\hbar\partial\psi/
\partial t$, as an {\em approximate} notion
 from the timeless Hamilton constraint equation
(the ``Wheeler-DeWitt equation") referred to
 above.\footnote{A detailed exposition on semiclassical
 gravity can be found in my contribution to
 {\em Canonical Gravity - From Classical to Quantum},
 edited by J. Ehlers and H. Friedrich (Springer,
 Berlin, 1994).} Secondly,
one may possibly get some insight, from semiclassical gravity,
into
 the interpretation of full quantum gravity.

I first want to focus on the recovery of standard time.
To this purpose it is convenient to write the Wheeler-DeWitt equation
in the form
\be \left(-\frac{\hbar^2}{2M}G_{abcd}\frac{\delta^2}
    {\delta h_{ab}\delta h_{cd}} +MV +{\cal H}_m\right)\Psi=0, \lb{1} \ee
where $M\equiv c^2/32\pi G$, $V=-2c^2\sqrt{h}(R-2\Lambda)$, and
${\cal H}_m$ is the matter Hamiltonian density.
 The semiclassical approximation to quantum
gravity is based on the assumption that $M$ in (1) is
large compared to the corresponding scales (masses, energies, \ldots)
in the matter Hamiltonian. One can then perform a Born-Oppenheimer
type of expansion scheme, in which the gravitational degrees
of freedom take over the role usually played by the nuclei,
and the remaining fields play the role of the electrons.

Writing the total wave functional in the form $\Psi\equiv
\exp(iS/\hbar)$, and expanding $S$ in powers of $M$,
$S=MS_0 +S_1+M^{-1}S_2+\ldots$, one finds, first, that $S_0$
obeys the {\em Hamilton-Jacobi equation for gravity}. Since this
equation is equivalent to the field equations of general
relativity, one may say that the notion of a classical spacetime
has been recovered at this order (to each solution $S_0$ of the
Hamilton-Jacobi equation corresponds a whole class of spacetimes
as ``trajectories of three-geometries" in configuration space).
One can construct a {\em particular} spacetime by first calculating
the geometrodynamical momentum from $S_0$, $\pi^{ab}\equiv
M\delta S_0/\delta h_{ab}$, and then finding from it the ``velocity"
$\dot{h}_{ab}$ by specifying lapse and shift function. Given
some ``initial" three-geometry, one can then integrate the equations
of motion to find a whole spacetime with a {\em definite} foliation
and a {\em definite} choice of coordinates on each member of the
foliation. If one writes the wave functional in the next order
in the form
\be \Psi\approx C[h_{ab}]\exp(iMS_0[h_{ab}]/\hbar)
     \chi[h_{ab},\phi], \lb{2} \ee
where $C$ is a real prefactor, and $\phi$ stands symbolically for
non-gravitational fields, the functional $\chi$ obeys an
approximate Schr\"odinger equation
\be i\hbar \int d^3x G_{abcd}\frac{\delta S_0}{\delta h_{ab}}
    \frac{\delta\chi}{\delta h_{cd}}
    \equiv i\hbar \frac{\partial\chi}{\partial t}=
    \int d^3x{\cal H}_m \chi. \lb{3} \ee
The ``WKB time" $t$ is defined on configuration space, but also
yields -- by the construction given above -- a time parameter
in each of the spacetimes which follow from a choice of
$S_0$. I emphasize that there is no ``spacetime problem" (see Ref.~1)
in the semiclassical approximation, since a specific foliation
of the spacetime has been chosen at the previous order of approximation.
Moreover, the fact that WKB time cannot be defined globally
(``global time problem") is of no importance in the semiclassical
approximation, since the Schr\"odinger equation (3) is expected
to hold only in restricted regions of superspace.

One might wonder what would happen if a {\em superposition}
of wave functionals of the form (2) is considered.  One might,
in particular, wish to consider a superposition of (2) with its
complex conjugate, which seems to be more natural in view of the real
nature of Eq. (1). In fact, due to the presence of interference terms,
there would be no definite spacetime, and one would
not be able to derive the Schr\"odinger equation
(3) from such a superposition. How, then,
is it possible to understand the emergence of a classical world?
The key observation in the resolution of this ``paradox"
is provided by the presence of a huge number of
degrees of freedom (such as gravitational waves and matter density
fluctuations) which couple differently to the $e^{iS_0}$ and
$e^{-iS_0}$ part of the superposition. Since only very few of these
degrees of freedom can be taken into account in a practical
observation, information about interference terms migrates into
correlations between the gravitational ``background part" and the
remaining ``reservoir" -- the terms in the superposition
{\em decohere} from one another.\footnote{C. Kiefer, Phys. Rev.
{\bf D} 47, 5414 (1993).}

It is important to note that, due to the Born-Oppenheimer
approximation, the unobservable degrees of freedom distinguish between
the {\em complex} components $e^{iS_0}$ and $e^{-iS_0}$ and {\em not}
between real components such as $\cos S_0$ and $\sin S_0$. This is
an important example for the mechanism of {\em symmetry
 breaking}\footnote{H. D. Zeh, in {\em Stochastic evolution
of quantum states in open systems and measurement processes}
(World Scientific, Singapore, 1994).} --
while the full Hamiltonian in (1) is invariant under complex conjugation,
the actual ``classical state" (2) is not. This is in close analogy
to molecular physics, where the Hamiltonian is invariant under
space reflection but the actual state can have a definite
chirality (this happens, e.g., in the case of sugar). In the same
way does time emerge from the timeless equation (1) by symmetry
breaking.

What happens with the concept of WKB time if one proceeds to the
next order in $M$? It is clear that now the phase of the functional
$\chi$ is involved in addition to $S_0$. In concreto, one defines
a ``corrected WKB time" $\tilde{t}$ by
\bea i\hbar\frac{\partial}{\partial\tilde{t}} & \equiv &
     i\hbar\int d^3x G_{abcd}\left(\frac{\delta S_0}
      {\delta h_{ab}}+ \frac{1}{M} \langle\chi\vert
      \frac{\delta\theta}{\delta h_{ab}}\chi\rangle\right)
      \frac{\delta}{\delta h_{cd}} \nonumber\\
     & = & i\hbar\int d^3x G_{abcd}\left(\frac{\delta S_0}
      {\delta h_{ab}}- \frac{1}{M} \langle\chi\vert
      {\cal H}_m \chi\rangle\right)
      \frac{\delta}{\delta h_{cd}}, \lb{4} \eea
 where $\theta$ is the phase of $\chi\equiv Re^{i\theta/\hbar}$.
 One recognizes in (4) the appearance of the ``back reaction
 term" $\langle{\cal H}_m\rangle$ which also appears in the corresponding
 modification of the Hamilton-Jacobi equation (and, consequently,
 in the Einstein field equations). Instead of (3) one finds at this order
 a Schr\"odinger equation with additional correction terms which are
 proportional to the gravitational constant. I emphasize that one
 contribution to these correction terms follows directly from
 the modification (4) of WKB time at this order (see, again,
 Ref.~2 for details). Similarly, one has to take into account
 the phase of the total wave functional up to order $M^{-k}$
 in the definition of WKB time, if one considers terms up to order
 $M^{-(k+1)}$ in the corrected Schr\"odinger equation. I should
 mention that one is able to extract physical predictions from
 these correction terms, such as energy shifts, and non-unitary
 contributions to black hole evaporation.

I will now focus on the second issue of my contribution: In what
sense can one extrapolate the notion of semiclassical time to full
quantum gravity? Motivated by the above considerations, a number of
authors\footnote{J. Greensite, Nucl. Phys. {\bf B 342}, 409 (1990);
T. Padmanabhan, Pramana {\bf 35}, L 199 (1990);
E. J. Squires, Phys. Lett. {\bf A 155}, 357 (1991).} have attempted
to {\em define} time in the full theory through the exact phase of
the total wave functional, which is supposed to satisfy the
Wheeler-DeWitt equation (1). This ``phase time", $t_p$, defines
a flow in configuration space which, of course, in general
has no connection with any classical spacetime. If one performs
a coordinate transformation in configuration space,
$(h_{ab},\phi)\to (t_p,y)$, where $y$ are ``comoving
coordinates", one can define expectation values according to
\be \langle A(y)\rangle \equiv
   \int{\cal D}\mu[y] \Psi^*(t_p,y) A\Psi(t_p,y) \lb{5} \ee
and show that $\langle A\rangle$ satisfies an Ehrenfest equation
$i\hbar \partial\langle A\rangle/\partial t_p
=\langle[A,\tilde{H}]\rangle$, where $\tilde{H}$ is an effective
Hamiltonian whose precise form, however, has to be determined
from a given solution to the Wheeler-DeWitt equation.

In the semiclassical approximation $t_p$ agrees, of course,
with WKB time. If the approximation scheme is continued to higher
orders, the modified WKB times are still in close analogy to
phase time. Such a connection is lost in generic situations
where the notion of semiclassical time is not available.

There are, however, some open problems with this approach to the
concept of time. Firstly, one cannot define phase time for a
real solution of the Wheeler-DeWitt equation. One might again
wish to invoke some kind of decoherence and define phase time in
various complex components of the total wave function, but this
is much less clear than in the Born-Oppenheimer approximation.
Secondly, in contrast to the semiclassical approximation, phase time
may depend upon the choice of the original spacetime foliation and thus
may give rise to the ``spacetime problem" (see Ref.~1). Thirdly,
the second Ehrenfest equation (the one with respect to momentum)
does not hold generally, but only if the phase obeys some additional
condition. Finally, time would not be a concept independent of
the chosen solution to the Wheeler-DeWitt equation. The viability
of this approach thus remains unclear.

I want to conclude with a brief comment on the relation of WKB time
to ``intrinsic time." The latter notion refers to the fundamental equation
(1) itself which through its
local minus signs in the kinetic part distinguishes locally
one function
of the three-metric (the local scale part)
 as being ``timelike". In special regions
in configuration space (for example, in the case of
Friedmann universes) only one minus sign remains, and the Wheeler-DeWitt
equation becomes hyperbolic with respect to intrinsic time.
While intrinsic time may in fact coincide with WKB time
for parts of the evolution of a Friedmann Universe, they will not
coincide globally. This, then, has important consequences for
the understanding of the arrow of time in quantum cosmology.\footnote{H.
D. Zeh, {\em The physical basis of the direction of time} (Springer,
Berlin, 1992).}

\end{document}